\documentclass{PoS}

\usepackage{cite}

\def\beq{\begin{equation}}
\def\eeq{\end{equation}}
\def\beqa{\begin{eqnarray}}
\def\eeqa{\end{eqnarray}}

\title{Charged Higgs production with a $W$ boson or a top quark}

\ShortTitle{Charged Higgs production with a $W$ boson or a top quark}

\author{\speaker{Nikolaos Kidonakis}\thanks{This material is based upon work supported by the National Science Foundation under Grant No. PHY 1519606.}\\
Department of Physics, Kennesaw State University, USA\\
        E-mail: \email{nkidonak@kennesaw.edu}}

\abstract{I present theoretical results for charged Higgs production in association with a $W$ boson or a top quark at the LHC. I calculate higher-order threshold corrections and show that they are very significant. I present detailed results for total cross sections as well as transverse-momentum and rapidity distributions for various LHC energies.}

\FullConference{The European Physical Society Conference on High Energy Physics\\
                 5-12 July 2017\\
                 Venice, Italy}

\begin{document}

\section{Introduction}

A charged Higgs discovery would be an unmistakable sign of new physics. Thus, the search for charged Higgs bosons is a significant part of current collider programs. Charged Higgs bosons appear in 2-Higgs doublet models, such as the MSSM, and the LHC has good potential for discovery of such particles via a variety of production modes. I discuss two important production processes: the associated production of a charged Higgs with a top quark, via the partonic process $bg \rightarrow t H^-$, and the associated production of a charged Higgs with a $W$ boson, via the partonic process $b{\bar b} \rightarrow H^- W^+$.

I show that higher-order corrections are significant for both processes. 
Given the very massive final states, soft-gluon corrections are important. 
I calculate these soft-gluon corrections through NNLO. In Section 2, I present 
approximate NNLO (aNNLO) results for total cross sections and top-quark differential distributions in $tH^-$ production. In Section 3, I present aNNLO results for charged-Higgs differential distributions in $H^- W^+$ production. 

\section{$tH^-$ production}

The top quark is the heaviest known elementary particle and it has unique properties, such as decay before hadronization. 
The lowest-order cross section for the process $bg \rightarrow t H^-$ 
is proportional to $\alpha \alpha_s (m_b^2\tan^2 \beta+m_t^2 \cot^2 \beta)$
where $\tan \beta=v_2/v_1$ is the ratio of the vevs of the two Higgs doublets.
NLO corrections for this process were calculated in Ref. \cite{tHcorr}.
It has long been known that soft-gluon corrections 
are important for this process \cite{NKcH,NKtWH,NKtH} as well as for related 
top-quark processes \cite{NKtt}.

For the process
$b(p_b) + g(p_g) \longrightarrow t(p_t)+H^-(p_H)$
we define $s=(p_b+p_g)^2$, $t=(p_b-p_t)^2$, $u=(p_g-p_t)^2$
and $s_4=s+t+u-m_t^2-m_H^2$.
At partonic threshold $s_4 \rightarrow 0$.

Soft-gluon corrections  appear as $\left[\frac{\ln^k(s_4/m_H^2)}{s_4}\right]_+$
where, for the order $\alpha_s^n$ corrections, $k \le 2n-1$.
We resum these soft corrections for the double-differential cross section.
At NLL accuracy this resummation was presented in \cite{NKcH}. 
At NNLL accuracy we need two-loop soft anomalous dimensions \cite{2loop,NKtWH}.
The resummed cross section can be expanded to NNLO using general techniques \cite{aNNLO}; aNNLO results from NNLL resummation for this process were first presented in Ref. \cite{NKtWH}. More recently, aNNLO total cross sections and top-quark transverse-momentum ($p_T$) and rapidity distributions in $tH^-$ production were presented in Ref. \cite{NKtH}.

To derive soft-gluon resummation, we first take moments of the partonic cross section with moment variable $N$:
${\hat \sigma}(N)=\int (ds_4/s) \; e^{-N s_4/s} {\hat \sigma}(s_4)$.  
Then, the moment-space factorized expression for the cross section in 4-$\epsilon$ dimensions is:
\beq
{\hat \sigma}^{bg \rightarrow tH^-}(N,\epsilon)= 
\left( \prod_{i=b,g} J_i\left (N,\mu,\epsilon \right) \right)
H^{bg \rightarrow tH^-} \left(\alpha_s(\mu)\right)\; 
S^{bg \rightarrow tH^-} \left(\frac{m_H}{N \mu},\alpha_s(\mu) \right) \, ,
\eeq
where $\mu$ is the scale, $J_i$ are jet functions, $H^{bg \rightarrow tH^-}$ is the hard-scattering function, and $S^{bg\rightarrow tH^-}$ is the soft function \cite{NKtH}. 

The soft function $S^{bg\rightarrow tH^-}$ satisfies the renormalization group equation
\beq
\left(\mu \frac{\partial}{\partial \mu}
+\beta(g_s, \epsilon)\frac{\partial}{\partial g_s}\right)\,S^{bg \rightarrow tH^-}
=- 2 \, S^{bg \rightarrow tH^-} \, \Gamma_S^{bg \rightarrow tH^-} \, .
\eeq
The soft anomalous dimension $\Gamma_S^{bg \rightarrow tH^-}$ is calculated at two loops \cite{NKtWH} and it controls the evolution of $S^{bg\rightarrow tH^-}$, resulting in the exponentiation of logarithms of $N$ \cite{NKtWH,NKtH}.

The aNNLO soft-gluon corrections from the resummation are:
\beq
\frac{d^2{\hat \sigma}_{\rm aNNLO}^{(2) \, bg \rightarrow tH^-}}{dt \, du}=F_{\rm LO}^{bg \rightarrow tH^-} \frac{\alpha_s^2}{\pi^2} 
\sum_{k=0}^3 C_k^{(2)} \left[\frac{\ln^k(s_4/m_H^2)}{s_4}\right]_+ \, ,
\eeq
where $F_{LO}^{bg \rightarrow tH^-}$ is the leading-order term and the $C_k^{(2)}$ coefficients are given in \cite{NKtH}.

\begin{figure}
\begin{center}
\includegraphics[width=74mm]{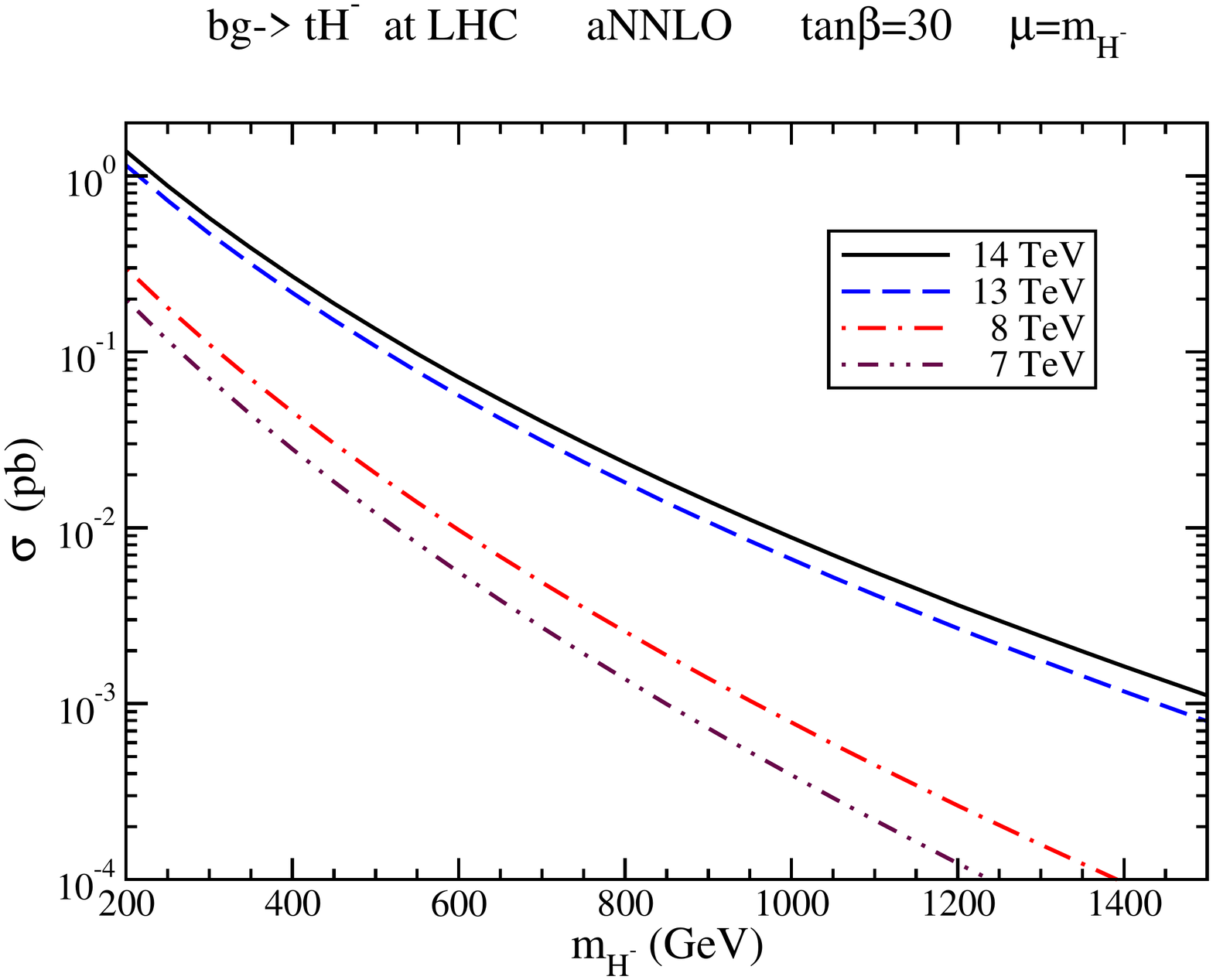}
\includegraphics[width=74mm]{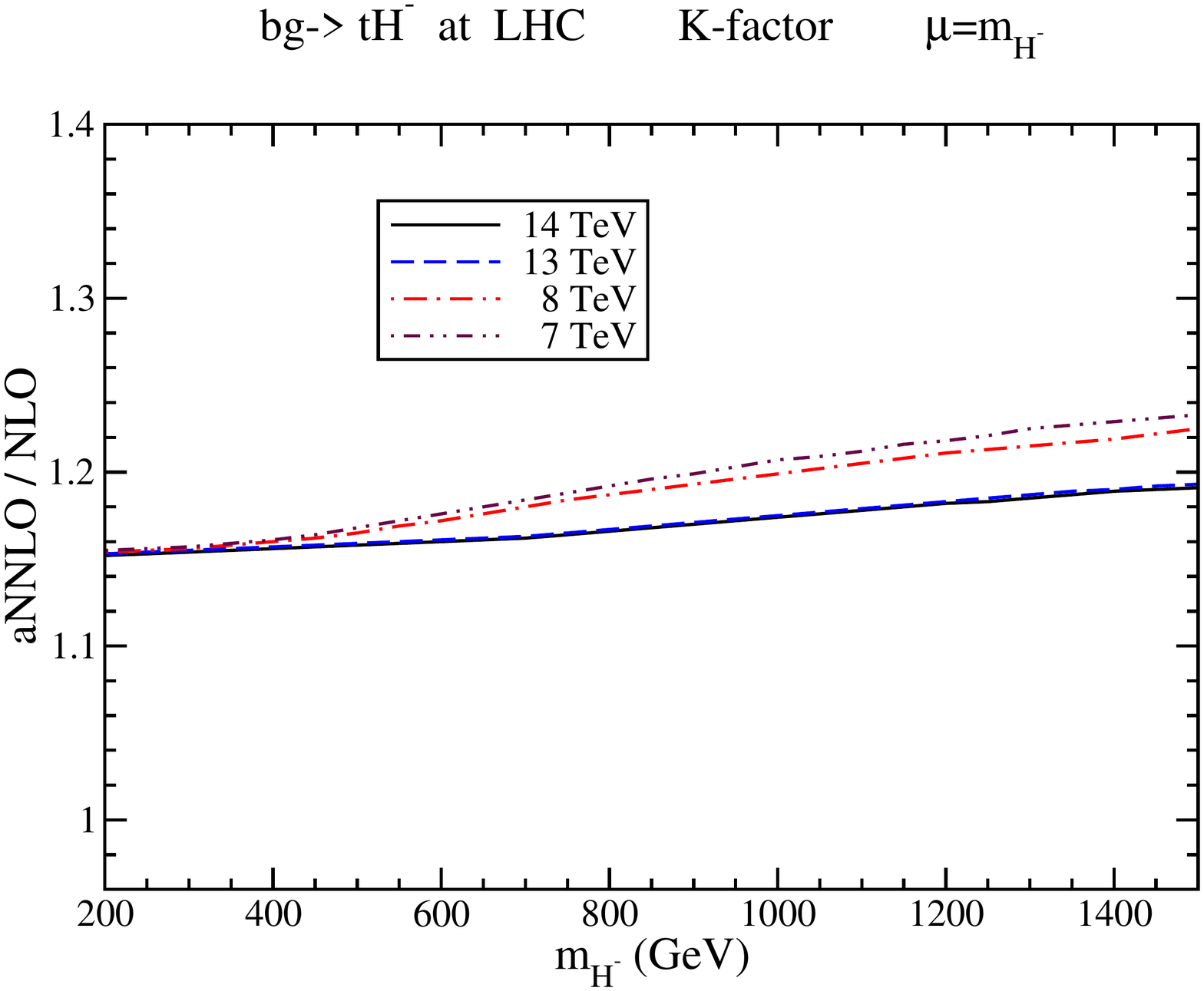}
\caption{Total cross sections with $\tan\beta=30$ (left), and $K$ factors (right) for $tH^-$ production.}
\label{tHaNNLO}
\end{center}
\end{figure}

We begin with the total cross sections for $tH^-$ production using MMHT2014 NNLO pdf \cite{MMHT2014}. In the left plot of Fig. \ref{tHaNNLO} we show the total cross sections at 7, 8, 13, and 14 TeV LHC energies with $\tan\beta=30$ for a wide range of charged Higgs masses. The aNNLO/NLO $K$-factors are shown in the plot on the right. The aNNLO corrections are large and can reach 20\% or more for high charged Higgs masses. The corrections are larger for lower LHC energies, closer to threshold. 

\begin{figure}
\begin{center}
\includegraphics[width=75mm]{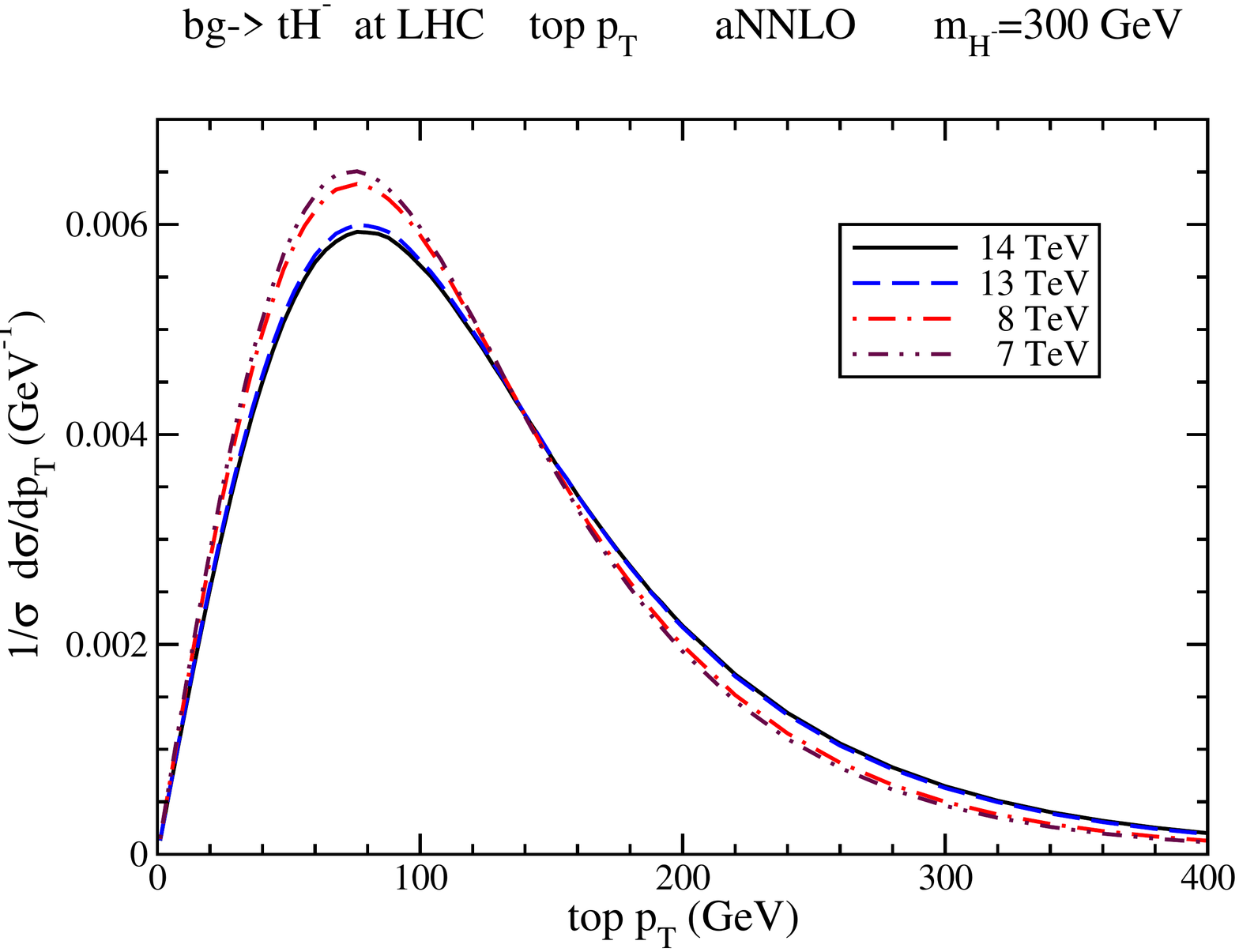}
\includegraphics[width=75mm]{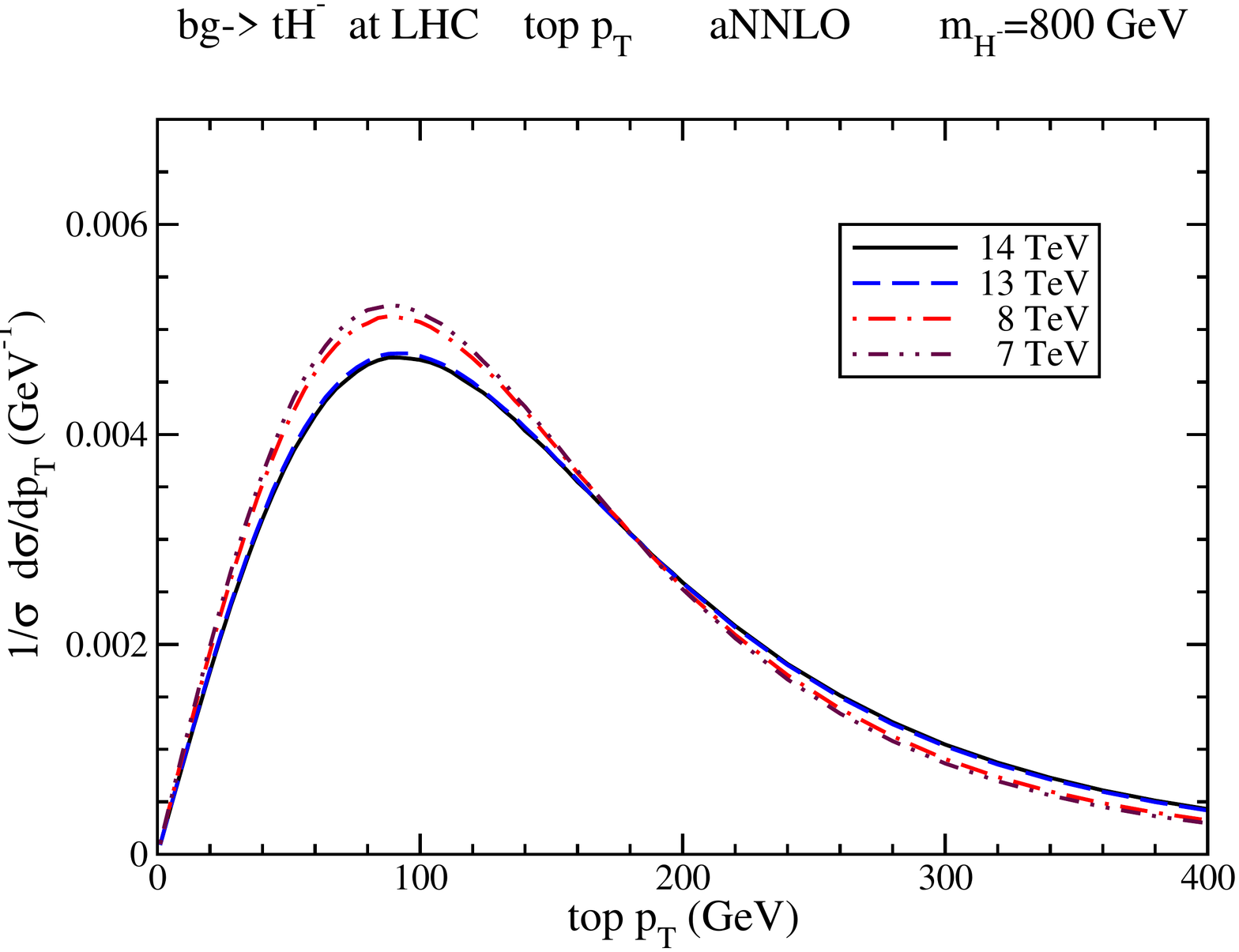}
\caption{Normalized top-quark $p_T$ distributions for $tH^-$ production with 
$m_H=300$ GeV (left) and 800 GeV (right).}
\label{pttHaNNLO}
\end{center}
\end{figure}

We continue with the transverse momentum distributions of the top quark in $tH^-$ production. In Fig. \ref{pttHaNNLO} we show the aNNLO normalized top-quark $p_T$ distributions, $(1/\sigma) d\sigma/dp_T$. The left plot shows results for a charged Higgs mass of 300 GeV while the plot on the right is for 800 GeV. We find significant corrections at all LHC energies. The shapes of the normalized distributions change as the energy is increased: at lower LHC energies the peak is higher and it appears at lower $p_T$, while at larger $p_T$ the curves from higher energies are larger.

\begin{figure}
\begin{center}
\includegraphics[width=75mm]{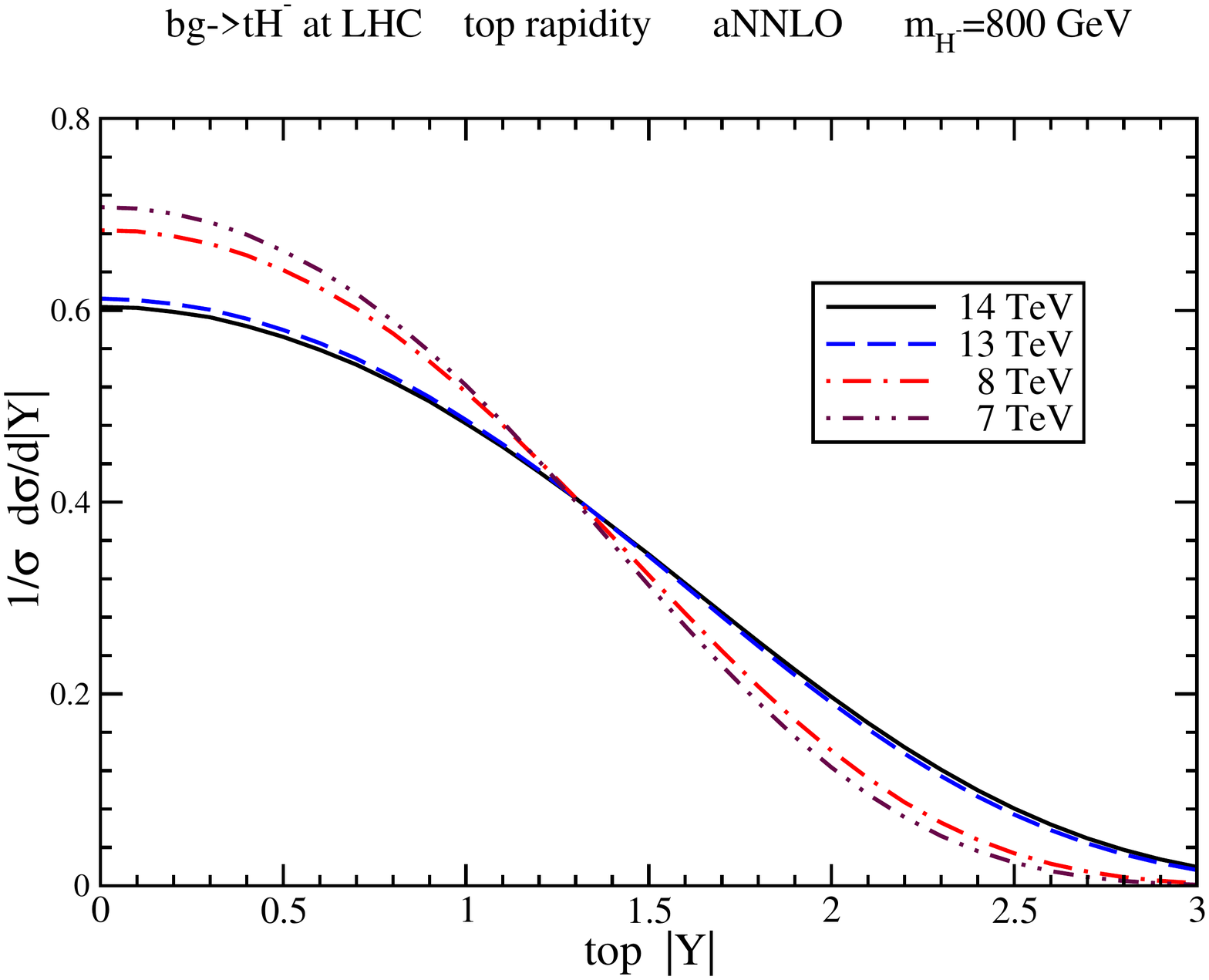}
\includegraphics[width=75mm]{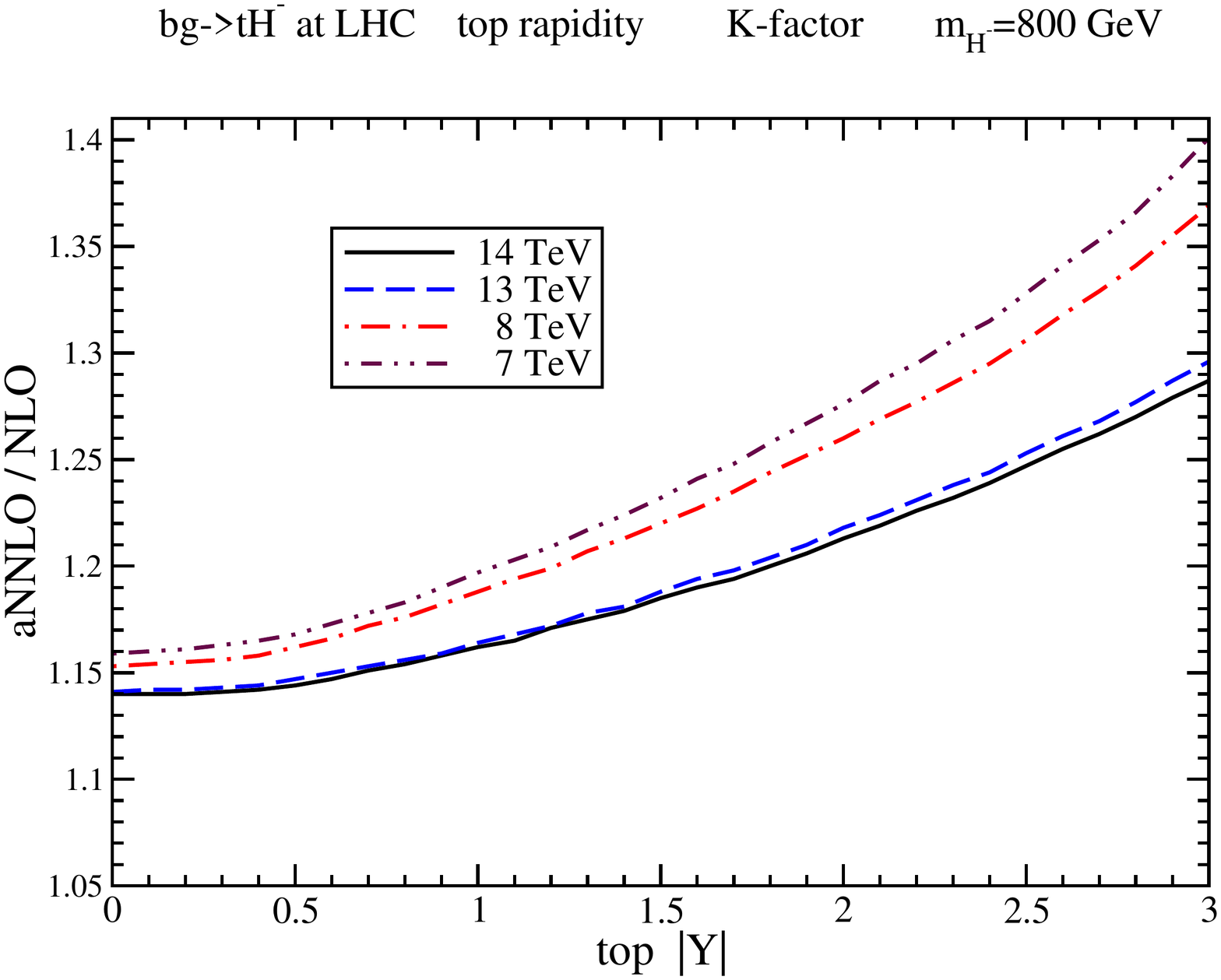}
\caption{Normalized top-quark rapidity distributions (left), and $K$ factors (right) for $tH^-$ production with $m_H=800$ GeV.}
\label{ytHaNNLO}
\end{center}
\end{figure}

In Fig. \ref{ytHaNNLO} we show results for the top-quark rapidity distributions in $tH^-$ production with a charged Higgs mass of 800 GeV. The left plot shows the aNNLO normalized top-quark rapidity distributions, $(1/\sigma) d\sigma/d|Y|$.
Again, the shape of the normalized distributions changes with energy, with the curves at higher LHC energies being higher at larger rapidities. The plot on the right shows the aNNLO/NLO $K$ factors. We find significant corrections, especially at large rapidities. The $K$ factors are larger at lower LHC energies, as expected.

\section{$H^-W^+$ production}

We continue with the associated production of a charged Higgs boson with a $W$ boson, via the process \cite{DHKR}
\beq
b(p_1)\, + \, {\bar b}\, (p_2) \rightarrow H^-(p_3)\, + W^+(p_4) \, .
\eeq
We define $s=(p_1+p_2)^2$, $t=(p_1-p_3)^2$, $u=(p_2-p_3)^2$,  
and $s_4=s+t+u-m_H^2-m_W^2$.
At partonic threshold $s_4 \rightarrow 0$. 
Soft-gluon corrections are important for this process \cite{bbHW}. 
In addition to the soft-gluon terms, here we also calculate leading terms of purely collinear origin, $\ln^k(s_4/m_H^2)$ \cite{bbHW}.

The NNLO collinear and soft-gluon corrections are \cite{bbHW}
\beq
\frac{d^2{\hat{\sigma}}_{\rm aNNLO}^{(2) \, b{\bar b} \rightarrow H^- W^+}}{dt \, du}= F_{LO}^{b{\bar b} \rightarrow H^- W^+} 
\frac{\alpha_s^2}{\pi^2}
\left\{-C_3^{(2)} \frac{1}{m_H^2} \ln^3\left(\frac{s_4}{m_H^2}\right) 
+\sum_{k=0}^3 C_k^{(2)} \left[\frac{\ln^k(s_4/m_H^2)}{s_4}\right]_+ \right\} \, .
\eeq

\begin{figure}
\begin{center}
\includegraphics[width=75mm]{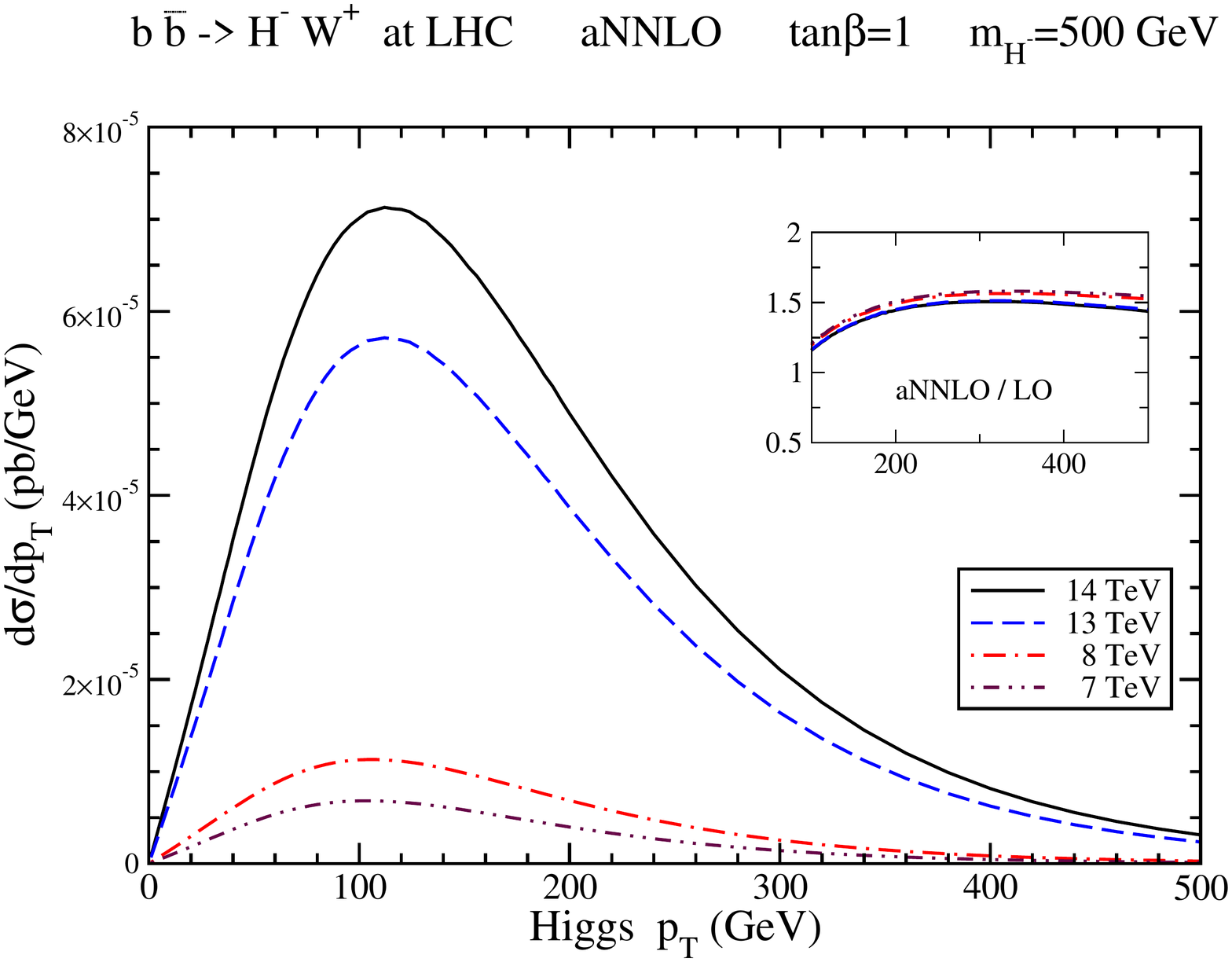}
\includegraphics[width=75mm]{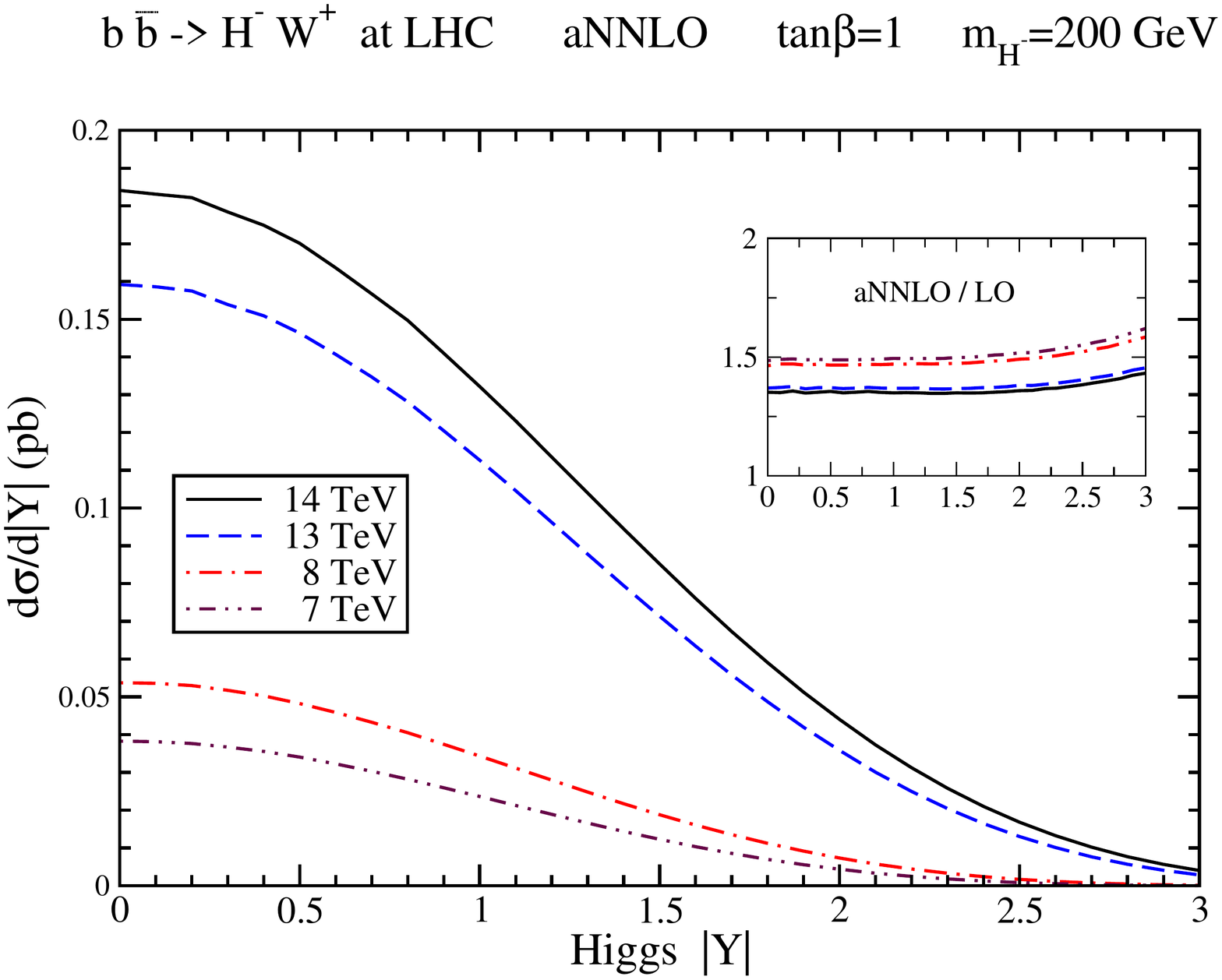}
\caption{Charged Higgs $p_T$ distributions with $m_H=500$ GeV (left), and rapidity distributions with $m_H=200$ GeV (right), for $H^-W^+$ production with $\tan\beta=1$.}
\label{HWaNNLO}
\end{center}
\end{figure}

In Fig. \ref{HWaNNLO} we show the aNNLO charged Higgs $p_T$ and rapidity distributions in $H^-W^+$ production with $\tan\beta=1$, using MMHT2014 NNLO pdf \cite{MMHT2014}. We find significant corrections for both distributions, as shown in the inset plots. We note that the results using the CT14 NNLO pdf \cite{CT14} are very similar.


\begin{thebibliography}{99}

\bibitem{tHcorr}
A. Belyaev {\it et al.}, 
Phys. Rev. D {\bf 65}, 031701 (2002); JHEP 06 (2002) 059; 
S.-H. Zhu, Phys. Rev. D {\bf 67}, 075006 (2003);
G.-P. Gao {\it et al.}, Phys. Rev. D {\bf 66}, 015007 (2002); 
T. Plehn, Phys. Rev. D {\bf 67}, 014018 (2003);
E.L. Berger {\it et al.}, Phys. Rev. D {\bf 71}, 115012 (2005);
W. Peng {\it et al.}, Phys. Rev. D {\bf 73}, 015012 (2006); 
S. Dittmaier {\it et al.}, Phys. Rev. D {\bf 83}, 055005 (2011); 
M. Flechl {\it et al.}, Phys. Rev. D {\bf 91}, 075015 (2015); 
C. Degrande {\it et al.}, JHEP 10 (2015) 145.

\bibitem{NKcH}
N. Kidonakis, JHEP 05 (2005) 011 [hep-ph/0412422]. 

\bibitem{NKtWH}
N. Kidonakis, Phys. Rev. D {\bf 82}, 054018 (2010) [arXiv:1005.4451 [hep-ph]].

\bibitem{NKtH}
N. Kidonakis, Phys. Rev. D {\bf 94}, 014010 (2016) [arXiv:1605.00622 [hep-ph]].

\bibitem{NKtt}
N. Kidonakis, Phys. Rev. D {\bf 90}, 014006 (2014) [arXiv:1405.7046 [hep-ph]]; 
Phys. Rev. D {\bf 91}, 031501 (2015) [arXiv:1411.2633 [hep-ph]]; 
Phys. Rev. D {\bf 91}, 071502 (2015) [arXiv:1501.01581 [hep-ph]];
Phys. Rev. D {\bf 96}, 034014 (2017) [arXiv:1612.06426 [hep-ph]].

\bibitem{2loop}
N. Kidonakis, Phys. Rev. Lett. {\bf 102}, 232003 (2009) [arXiv:0903.2561 [hep-ph]].

\bibitem{aNNLO}
N. Kidonakis, Int. J. Mod. Phys. A {\bf 19}, 1793 (2004) [hep-ph/0303186]; 
Mod. Phys. Lett. A {\bf 19}, 405 (2004) [hep-ph/0401147].

\bibitem{MMHT2014}
L.A. Harland-Lang, A.D. Martin, P. Molytinski, and R.S. Thorne,   
Eur. Phys. J. C {\bf 75}, 204 (2015) [arXiv:1412.3989 [hep-ph]].

\bibitem{DHKR}
D.A. Dicus, J.L. Hewett, C. Kao, and T.G. Rizzo, Phys. Rev. D {\bf 40}, 787 (1989).

\bibitem{bbHW}
N. Kidonakis, arXiv:1704.08549 [hep-ph]. 

\bibitem{CT14}
S. Dulat, T.-J. Hou, J. Gao, M. Guzzi, J. Huston, P. Nadolsky, J. Pumplin, C. Schmidt, D. Stump, and C.-P. Yuan, Phys. Rev. D {\bf 93}, 033006 (2016) [arXiv:1506.07443 [hep-ph]].

\end{thebibliography}
\end{document}